\documentclass[10pt,twocolumn,letterpaper]{article}
\usepackage{iccv}
\usepackage{mathptmx} 
\usepackage{graphicx}
\usepackage{amsmath}
\usepackage{amssymb}
\usepackage{tikz}

\usepackage{amssymb}
\usepackage{url}            %
\usepackage{booktabs}       %
\usepackage{amsfonts}       %
\usepackage{amssymb}
\usepackage{amsmath}
\usepackage{nicefrac}       %
\usepackage{microtype}      %
\usepackage[singlelinecheck=off]{caption}

\usepackage{subcaption}

\usepackage[sort, nocompress]{cite}
\usepackage{graphicx} %
\usepackage{multirow}
\usepackage{setspace}
\usepackage{enumitem}
\usepackage{wrapfig}

\usepackage{xcolor}
\usepackage{colortbl}

\usetikzlibrary{shapes.misc, shapes.multipart, positioning}

\usepackage [latin1]{inputenc}
\usepackage[pagebackref=true,breaklinks=true,colorlinks,bookmarks=false]{hyperref}
\iccvfinalcopy
\def\iccvPaperID{7091} %

\ificcvfinal\fi

\long\def\RDZignore#1{}

\title{Deep survival analysis with longitudinal X-rays for COVID-19}

\author{Michelle Shu\textsuperscript{1} \hspace{0.3 cm}
Richard Strong Bowen\textsuperscript{1} \hspace{0.3 cm}
Charles Herrmann\textsuperscript{1} \hspace{0.3 cm} 
Gengmo Qi\textsuperscript{1} \\
Michele Santacatterina\textsuperscript{3} \hspace{0.3 cm}
Ramin Zabih\textsuperscript{1,2}\\
\textsuperscript{1} Cornell Tech \hspace{0.3 cm} \textsuperscript{2} Google Research \hspace{0.3 cm} \textsuperscript{3} George Washington University\\
{\tt\small \{michelleshu, rsb, cih, rdz\}@cs.cornell.edu, gq35@cornell.edu, msanta@email.gwu.edu}\\
}

\definecolor{yellow}{rgb}{1,1, 0.6}
\definecolor{orange}{rgb}{1, 0.8, 0.6}
\definecolor{red}{rgb}{1, 0.6, 0.6}
\definecolor{darkred}{rgb}{0.8, 0.0, 0.0}
\definecolor{blue}{rgb}{0, 0, 1.0}
\definecolor{green}{rgb}{0, 1.0, 0}
\definecolor{darkgreen}{rgb}{0, 0.4, 0}
\definecolor{darkorange}{rgb}{1, 0.6, 0}

\definecolor{Aero Blue}{rgb}{1,1, 0.6}
\definecolor{Magic Mint}{rgb}{1, 0.8, 0.6}
\definecolor{Pearl Aqua}{rgb}{1, 0.6, 0.6}

\begin{document}

\maketitle
\ificcvfinal\thispagestyle{empty}\fi

\begin{abstract}
Time-to-event analysis is an important statistical tool for
allocating clinical resources such as ICU beds. 
However, classical 
techniques like the Cox model cannot directly incorporate images due to their
high dimensionality. We propose a deep learning approach that naturally incorporates multiple, time-dependent imaging studies as
well as non-imaging data into time-to-event analysis. Our techniques are benchmarked on a clinical dataset of 1,894 COVID-19
patients, and show that image sequences significantly improve predictions.
For example, classical time-to-event methods produce a concordance error of around 30-40\% for predicting hospital admission, while our error is 25\% without images and 20\% with multiple X-rays included. 
Ablation studies suggest that our models are not learning spurious features such as scanner artifacts and that models which  use multiple images tend to perform better than those which only use one.
While our focus and evaluation is on COVID-19, the methods we develop are broadly applicable.
\end{abstract}

\section{Introduction}

COVID-19 is the most significant global health emergency in recent memory,
with hundreds of thousands dead and widespread economic disruption.
There is growing evidence that imaging is useful for the diagnosis and
management of COVID-19 \cite{Caruso2020,Tabata2020}.  Clinicians use radiology
imaging to assess structural information which cannot be assessed with
laboratory tests or physical examination.  In COVID-19, chest imaging adds a
high-dimensional assessment of the degree of pulmonary involvement of the
disease.  It allows clinicians to rule out other conditions which might
contribute to the patient's presentation such as lobular pneumonia and
pneumothorax and to assess the patient for comorbidities such as heart failure,
emphysema, and coronary artery disease.  Some researchers have already found
that imaging features predict mortality in COVID-19 \cite{Toussie2020}.

In this paper we address the challenge of predicting the time course of COVID-19 patient outcomes; for example, the probability that a specific patient will need an ICU bed in the next few days following hospital admission.
Classical statistical techniques for time-to-event analysis (sometimes referred to as survival analysis) are widely used, but struggle with incorporating images due to their high dimensionality. 

We begin with an overview of time-to-event analysis and a discussion of the challenges that images and COVID-19 present. %
Our deep learning approach is presented in section~\ref{sec:approach}, followed by a review of related work in section~\ref{sec:related}.
We describe our clinical dataset and some implementation details, including the baseline, in section~\ref{sec:setup}.
Experimental results are given in section~\ref{sec:data}, with additional data and analysis in the supplemental material.

\section{Time-to-event analysis}
\label{sec:TTE}

{Time-to-event analysis} techniques \cite{kleinbaum2010survival} predict the 
probability of an outcome event occurring before a specific time, while accounting 
for {right-censored} (incompletely observed) data.  
Right-censoring happens when the event under study may not be observed within the relevant time period.  
In the clinical setting, these methods can predict a patient's probability of 
undergoing an event in a particular time interval as a function of their features.
In our dataset for instance, when predicting if a hospitalized COVID-19 patient 
will be admitted to the ICU, right-censoring happens when, as of today, the patient 
has not been admitted.

Time-to-event analysis focuses on three interrelated quantities: (1) the
{hazard function}, the rate of an event at time \textit{t} given that
the event did not occur before time \textit{t}, which is not affected by
right-censoring \cite{beyersmann2011competing}; (2) the {cumulative
  hazard function}, the integral of the hazard function between 0 and time
\textit{t}; and (3) the {survival function}, a decreasing function that
provides the probability that a patient will not experience the event of
interest beyond any specified time \textit{t} and is expressed as the
exponential of the negative of the cumulative hazard function.

While the hazard function is not a probability, $h(t)$ can be viewed as the
probability of the event occurring in a small interval around $t$ given that
the event did not occur before $t$.  For clinical purposes, once we have
estimated the hazard function we can then compute the probability of an event
occurring during a specific time interval, e.g. ICU admission in the 72 hours after hospitalization.

\subsection{Cox proportional hazards model}

We model the hazard function using
the most popular model, the Cox model
\cite{cox1972regression}, defined as
  \begin{equation}
     h(t|x) = h_0(t) \exp{\left[ r( x(t) ) \right]}.
     \label{eq:hazard_function}
 \end{equation}
 
\noindent
Here $t$ is the time, $x(t)$ is the set of features, $h_0(t)$ is the baseline
hazard, the hazard of the specific event under study shared by all patients at
time $t$, and $r(x(t))$ is the \textit{risk function}, which describes the relationship between a
patient's features $x(t)$ and the hazard of experiencing an event. 
Note that $h_0(t)$ only depends on time and not on features.

The Cox model has several advantages. First, it has no distributional
assumptions, so its results will closely approximate the results for the
correct model \cite{kleinbaum2012cox}. Second, even though the baseline hazard
is assumed to be ``unknown'' and left unspecified, under minimal 
assumptions, the hazard, the cumulative hazard and the survival functions can
be directly determined. These can then be used for predicting the probability
of an event occurring before the observing time $t$\footnote{Another
  statistical technique used to predict probabilities of binary events is
  logistic regression \cite{kleinbaum2002logistic}. Although widely used,
  logistic regression ignores the information about the time to an event and
  the censoring time, while time-to-event techniques fully exploit this important
  information.}  \cite{kleinbaum2012cox}.

The Cox model assumes that \textit{time-fixed} features (features that do not
change over time) have a {linear} multiplicative effect on the hazard function
and that the hazard ratio is constant over time. This is known as the proportional
hazards (PH) assumption \cite{cox1972regression}.  In our task, this means that,
for example, patients with a low income have a higher (or lower) hazard of
dying compared with patients with a high income and this ratio is constant
over time. 
Note that this assumption is only needed for time-fixed features and successful strategies can be easily implemented to detect and
  overcome its violation. Examples include  the graphical approach of log-log plots for
  detection, as well as adding into the Cox model an interaction between the
  non-proportional time-fixed feature and time for overcoming its violation
  \cite{kleinbaum2012extension,kleinbaum2012evaluating}.

\subsection{Images present challenges}

The Cox model is a mainstay of time-to-event analysis, and has been extended to
deal with complex scenarios~\cite{faraggi1995neural,tibshirani1997lasso,katzman2018deepsurv,lee2018deephit,hao2018cox,leeDynamicDeepHitDeepLearning2020}.
However, there are two features of our task that require us to go beyond the
state of the art. First, images pose a significant challenge due to their high 
dimensionality. Second, the time course of COVID-19 involves multiple interrelated 
events that cannot be predicted independently.

While there is compelling evidence that imaging studies are helpful in the
diagnosis and management of COVID-19 \cite{Caruso2020,Tabata2020,Toussie2020},
images present significant challenges.  The amount of data in a single imaging
study is orders of magnitude larger than the data available from other
sources; a single medical image can easily be hundreds of
megabytes.\footnote{While CT is becoming increasingly available, early in
  the pandemic imaging was primarily chest X-ray (CXR).}  However, the Cox proportional
hazards model cannot directly handle images as features due to their high
dimensionality. {As \cite{simon2011regularization} reports, such inputs lead
  to degenerate behavior.  It would of course be possible to learn a feature
  from an imaging study, for example a rating of disease severity on a 3-point
  scale. Such an approach would severely and unnecessarily limit what can be
  learned from the images, which is a particularly poor choice for a novel
  disease.}

As mentioned, the COVID-19 disease process involves competing and interrelated events.
A straightforward application of time-to-event analysis would predict these
events independently. This could easily lead to incoherent and
self-contradictory predictions (for example, predicting that ICU discharge will almost certainly
happen before admission to the ICU).

\section{Our approach}
\label{sec:approach}

{Our main goal is to predict the probability of experiencing death, ICU
  admission, ICU discharge, hospital admission, and hospital discharge before the observing time
  $t$, as a function of patient's features}. To do so, we assume nonlinear
proportional hazards \cite{katzman2018deepsurv} for the time-fixed features.\footnote{This
  assumption is only needed for time-fixed features. Time-dependent features
  already depend on time, making the hazard also depending on time.}
This assumption relaxes the more strict assumption
of {linear} proportional hazards of the classical Cox model. In our
analyses this means that, for instance, the hazard of dying among older
patients at baseline increases non-linearly compared with younger patients at
baseline and this ratio is constant over time~\cite{kleinbaum2012cox}. This
assumption has been already used in a variety of state of the art deep
learning time-to-event techniques
\cite{faraggi1995neural,zhu2016deep,katzman2018deepsurv}. 
We also made the common assumption of non-informative censoring  \cite{kalbfleisch2011statistical} which states that after conditioning on observable features, censoring and the risk for an event of interest are independent, \textit{i.e.},  the censoring mechanism does not depend on the unobserved features.

To estimate the hazard function introduced in Eq. \eqref{eq:hazard_function}
we compute two components, the baseline hazard $h_0(t)$ that only depends
on the time $t$ and the risk function $r(x(t))$ that only depends on features
$x(t)$. Once the hazard function is estimated, the cumulative hazard function
and the survival function can be easily derived \cite{kleinbaum2010survival};
these are then used to predict the patient's probability of undergoing an event
before time $t$. In other words, for each patient's set of features $x(t)$, we
can predict the probability of an event happening before the observed time
$t$. The baseline hazard does not depend on the features $x(t)$ and therefore,
it can be easily computed by using classical estimators. We
used the one presented in Eq. 4.34 of
\cite{kalbfleisch2011statistical}. The risk function, however, depends on
time-fixed and time-dependent image and non-image features.  To estimate the
risk function $r(x(t))$ while taking into account these challenging types of
features, we developed novel deep learning techniques. 

\subsection{Architecture}

\begin{figure}
    \centering
    \includegraphics[width=\linewidth]{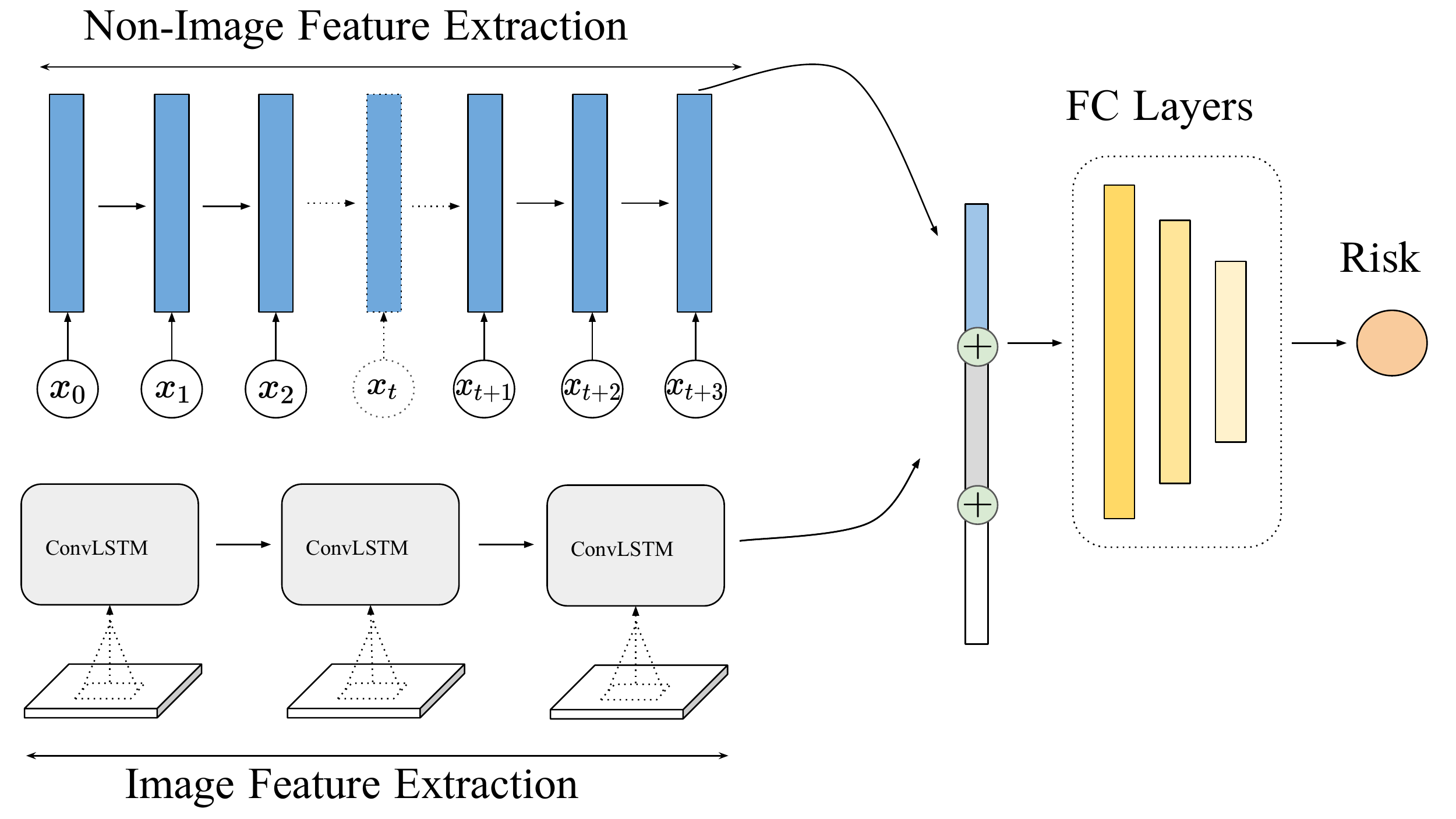}
    \caption{Our proposed architecture that handles time-fixed data, longitudinal non-image data and longitudinal images.}\label{fig:cnn2}
\end{figure}

To incorporate time-dependent imaging studies, time-dependent non-image data, and time-fixed variables, our proposed architecture has three components. First, we use a convolutional LSTM (ConvLSTM)\cite{ShiLSTM15} and an RNN-LSTM to extract time-dependent image features and time-dependent non-image features respectively. Then, we concatenate ($\oplus$) the features extracted from the networks with time-fixed variables mapped to its corresponding embedding space, and passed the concatenated vector through a set of fully connected layers (FC Layers) to predict the risk function (Risk). The architecture is shown in figure~\ref{fig:cnn2}.

\subsection{Loss function}
Computed on the hazard function, the Cox's partial likelihood loss has been successfully used
in recent state-of-the-art deep learning techniques
\cite{zhu2016deep,katzman2018deepsurv}. This likelihood function, however, is unsuitable, since it only applies to continuous time data where no two events occur at the same time. In the case of discrete time data, ties may occur and all possible orders of these tied events should be considered. Therefore, we adopt Efron's approximation for handling ties, which is a computationally efficient estimation on the original Cox's partial likelihood when ties are present\cite{efron1977efficiency}. 

Specifically, the loss function is as
follows:
  \begin{multline}
    L =
     \sum_{i} \Bigl[ R_i -
     \sum_{w=0}^{d_i -1} \log (\sum_{j:T_j \geq t_i} e^{ r(x_j((t)) } - \frac{w}{d_i} U_i) \Bigr],\\
     \text{where}\quad
     R_i = \sum_{j \in K_i} r(x_j(t)), U_i = \sum_{j \in K_i} e^{r(x_j(t))}.
     \label{eq:hazard_function_gj}
  \end{multline}
\noindent
 Here $i$ denotes a unique time point and $T_j$ is the followup time for patient $j$. We define the event indicator $C_j$, where $C_j = 1$ if the patient $j$ experiences an event at followup time $T_j$, and $C_j = 0$ if censored. The risk estimate $r(x_j(t))$ is the output of our architectures for patient $j$. $K_i$ is the set of patients whose followup time $T_j = t_i$ and $C_j = 1$, and $d_i = |K_i|$. It is worth noticing that by construction, this loss function does not contain the baseline hazard $h_0(t)$, making its computation easier. 
 
 \subsection{Evaluation and inference}
 \label{sec:eval-and-infer}
 
 We use concordance error \cite{uno_c-statistics_2011} to compare performances of different models. This calculates the proportion of falsely ordered event pairs divided by the total number of possible evaluation pairs.
 
 Similar to the loss function, during inference time, our model estimates the baseline hazard $\Delta H_0(t_i)$ and the cumulative baseline hazard $H_0(t)$ function using the Efron estimator for handling ties:
 \begin{align}
    \Delta H_0(t_i) & = 
     \sum_{l=0}^{D_i -1} \frac{1}{\sum_{j:T_j \geq t_i} e^{ r(x_j) } - \frac{l}{D_i} \sum_{j \in K_i} e^{r(x_j)}},\\
    H_0(t) & = \sum_{t_i \leq t}  \Delta H_0(t_i)
  \end{align}

\subsection{Mini-batch SGD and stratified sampling}
\label{sec:mini-batch}

In the Cox partial likelihood function, we see that the formulation involves the risk predictions of all the patients whose followup time is longer than $T_i$. It is computationally costly and almost infeasible to optimize this loss when models use time-dependent images. Instead, we use mini-batch stochastic gradient descent: for each iteration, we sample a subset of patients and compute the loss on the subset. To closely mimic the data distribution of the original patient group and to keep the loss function in a stable range (the range of loss function correlates to the number of events in the patient group), we use stratified sampling to maintain the same ratio of non-censored and censored patients in each mini-batch.

\section{Related work}
\label{sec:related}

Time-to-event analysis is an important tool for many applications and plays a crucial role in healthcare. 
A wide variety of techniques have been developed \cite{kleinbaum2010survival}, notably including non-parametric methods such as Kaplan-Meier, parametric methods such as Weibull and Gompertz, and semi-parametric models, such as the Cox model. 
Powerful machine learning algorithms have also been adapted to this task; popular examples include Survival Support Vector Machines\cite{khan2008support} and Random Survival Forests\cite{breiman2001random}.
The Cox model, which directly estimates the hazard function, stands out as perhaps the most popular time-to-event analysis method.

\textbf{Deep Survival Analysis.} In recent years, deep learning has been used to extend the Cox model. DeepSurv\cite{katzman2018deepsurv} was one of the first to use deep learning to model the risk function $r(x(t))$ in the hazard function. They demonstrated strong performance in both linear and non-linear settings. 
There are also attempts to tackle the limitations of DeepSurv to only structured data. 
By using convolution neural networks, \cite{zhu2016deep, Zhu2017, zheng2018feature} models hazards on unstructured features such as images, which are much harder to incorporate due to their high dimensionality. 
Our architecture further extends the Cox model so that it takes both structured and unstructured longitudinal input.

\textbf{Unstructured Longitudinal Data.} Incorporating longitudinal data in time-to-event prediction has received increasing attention, typically using deep learning techniques. 
\cite{lee2018deephit} proposes DeepHit, a model that enables the possibility that the relationship between covariates and risks to change over time. \cite{jin2020deep, giunchiglia2018rnn,ren2019deep} have demonstrated the effectiveness of recurrent neural networks on longitudinal structured datasets. 
We appear to be the first to adopt recurrent neural network on longitudinal unstructured medical data, and the recurrent nature of our model is successful in capturing time dependency relationships within the data.

\textbf{End-to-End Training.} As mentioned in section~\ref{sec:mini-batch}, the negative partial likelihood function loss function is computationally expensive and almost infeasible (in terms of GPU memory) to compute when we have a large patient group with unstructured image data. 
\cite{peng2020predicting} adopted an alternative two-step training strategy which requires first training a feature extraction network on image data with expert labelling, then uses the extracted features as covariates of the Cox model. 
Using mini-batch sampling, our training process is end-to-end and  does not require any expert labelling.

\section{Experimental setup and clinical dataset}
\label{sec:setup}

Our techniques are designed for a clinical setting, where a combination of
time-dependent and time-independent patient features are available, including
imaging.  There are no publicly available datasets for COVID-19 that contain
this information, and patient privacy considerations make it unlikely that
such data will be available anytime soon.  The closest existing datasets for
COVID-19 focus primarily on images, and generally do not contain significant
additional information.  
The recent BIMCV-COVID19+ dataset
\cite{vaya2020bimcv} is an exception, and contains a limited amount of
information such as demographics and antibody test results, but falls far
short of the detailed clinical information that our methods are designed to exploit. 
Notably, it does not contain patient outcomes or lab values.

\subsection{Clinical dataset}

\begin{table*}
    \centering
    \captionsetup{justification=centering}
\begin{tabular}{|c|c|c|c|c|c|}
\hline
   & Admit & ICU Admit & ICU Discharge & Discharge & Death \\
\hline\hline
Event \# & 137 & 287  & 251   & 1171 & 290 \\
Censored \# & 99 & 224  &  150  & 219 & 427 \\
Event Removed \# & 1395 & 138  & 1   & 26 &  6\\
Censored Removed \# & 263 &  1245 & 1492    & 478  & 1171 \\
\hline\hline
Baseline Date & First available X-ray   & Admit   &  ICU  Admit &  Admit & Admit     \\
Start Date & Same as Baseline & 7 day Before  &  7 Day Before  & 7 Day Before & 7 Day Before\\
Cut-Off Day & 30 Days & 10 Days  &  10 Days  & 10 Days & 10 Days \\
\hline
\end{tabular}
    \caption{Details on the data distribution for each type of events.}
    \label{tab:data_distribution}
\vspace{0.5cm}
\begin{tabular}{|c|c|c|c|c|}
\hline
   & Age & Smoking & Pregnancy & Cancer\\
\hline\hline
Data Breakdown & IQR: 23  & Active Smoker: 70 & Yes: 7 & Solid: 71 \\
                      & Median: 64 & Former Smoker: 382  & No: 1705 & Liquid: 42 \\
                      &            & Non-Smoker: 1257  &   & No: 1257 \\
\hline                      
Missing Report & 29.8\% & 9.8\% & 9.6\% & 27.6\% \\
\hline
\end{tabular}
    \caption{Time-fixed features in our clinical dataset.}
    \label{tab:demographics}
    \vspace{-0.5cm}
\end{table*}

\begin{figure}
    \centering
    \includegraphics[width=\linewidth]{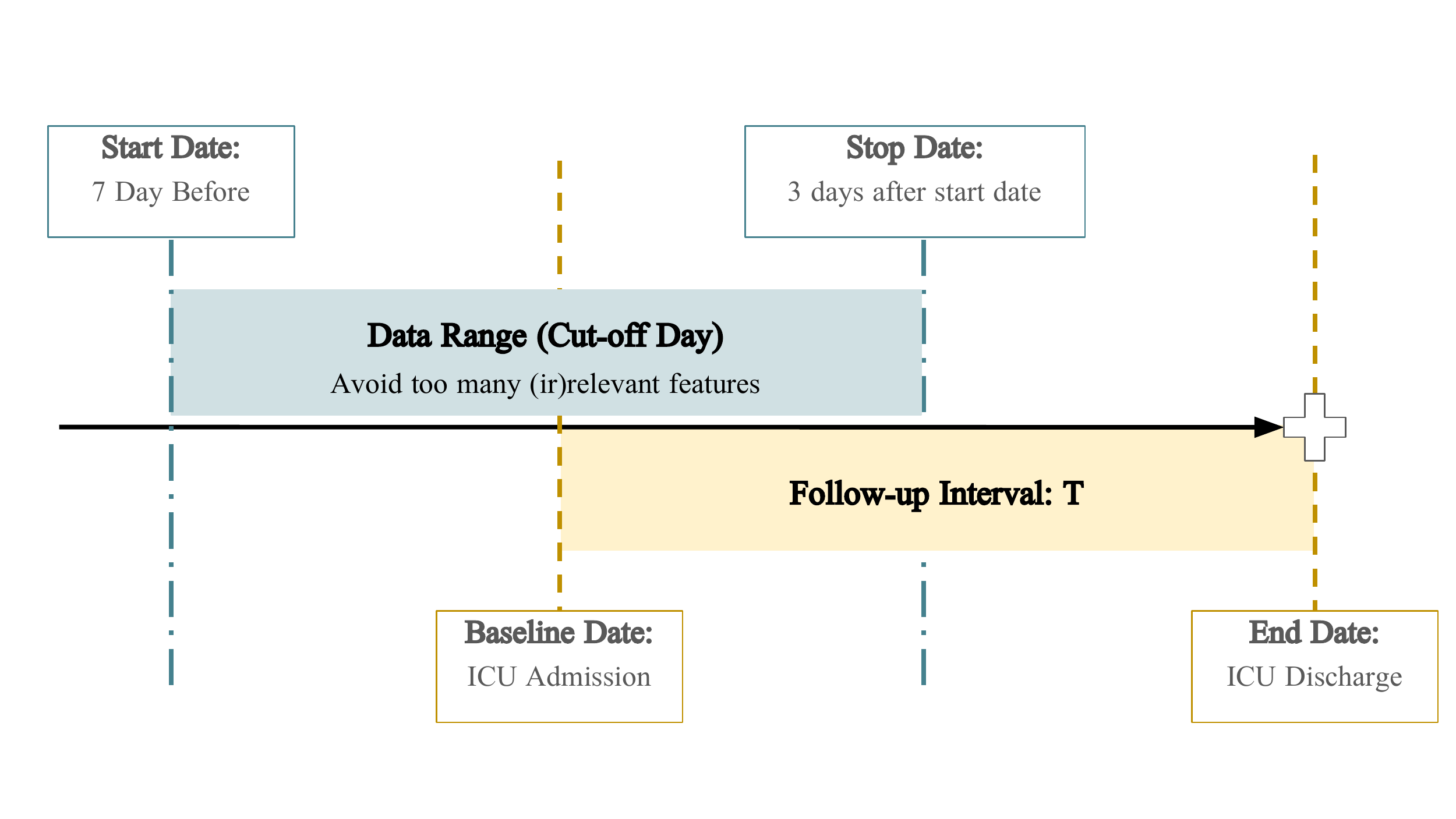}
    \caption{A sample timeline for ICU Discharge event prediction. The baseline date for ICU Discharge is ICU Admission. To avoid giving models too much information, we set the data range of this event to be 10 days. Any data before and after this range will be removed.}\label{fig:timeline}
    \vspace{-5mm}
\end{figure}

We are fortunate to have IRB-approved access to a large clinical dataset of 1,894 COVID-19 patients, longitudinally collected from 90 consecutive days early in the
pandemic.  
The dataset includes patient demographics, hourly-recorded
vital signs, treatment regimes and clinical notes. 
Crucially for our purposes there are a significant number of imaging studies  also available: over 14,000 studies, primarily chest X-ray (CXR) but also an increasing number of computed tomography (CT) exams.  
Outcomes are also available, including such important events as hospitalization admission and discharge, ICU admission and discharge, and mortality.
All patients tested COVID-19 positive by PCR.
Multiple X-rays, 4 types of time-fixed features and 29 types of time-dependent non-image features of patients during their stay are also available.
Time-fixed features are smoking and pregnancy status, active cancer and age at admission.

The data was provided to us by a study research coordinator.
Due to possible inconsistencies in the ways that events are recorded at their institution, there may be a small number of patients whose date of events was not included. 
In such a case, the patient's events occurred but were unobserved, which are right-censored data in a time-to-event framework.
Our approach to censored outcomes, discussed below, handles this situation straightforwardly.

We process the available data with the following pipeline:
 \begin{itemize}
 \item For each type of event, we select a baseline date for the Cox model based on advice from  clinicians, and generate a time interval for each patient. 
 The start date and end (followup) date of the interval is dependent on the baseline date and the type of events. We filter out any patient without an appropriate baseline date.
 \item We discretize the time interval generated for each qualified patient by day, and associate each time-dependent feature in this time interval with their corresponding days. We remove patients who do not have any X-rays taken during this interval.
 \item A cut-off day is selected for each type of event, and we remove any data collected beyond this cut-off day.
\end{itemize}

Table~\ref{tab:data_distribution} summarizes our choices of baseline date, start date and cut-off day for each type of event. For subjects with event outcomes, the date of the followup is exactly the date when the event occurred. For censored outcomes (including unrecorded deaths), we use the date of the most recent X-ray to be the followup date, where the study terminates without reaching a definite conclusion. Statistics on time-fixed features are provided in Table~\ref{tab:demographics}. For completeness, the set of time-dependent features can be found in the supplements.
 
In our dataset, longitudinal information is available until the patient reaches an absorbing state. 
This is not a realistic setting, since predictions are mostly likely needed at earlier stages, where only data close to the baseline date are available. 
However, our cut-off day procedure prevents simple cheating such as counting the number of days where data is available.
The cut-off also avoids other forms of cheating, for example if we go far enough beyond the baseline date, some lab results may start to give obvious clues.

Note that our model is recurrent and capable of analyzing input without any X-rays. 
However, we want to study the impact of images on time-to-event analysis, and wish to compare our model with another image-based survival analysis method -- DeepConvSurv \cite{zhu2016deep}.
As a result we we have filtered out patients who did not have any X-rays taken during the corresponding intervals. 

After obtaining the desired patient population for each event, we follow the standard train, validation, test split (60:20:20). We divide the population to train, validation and test set such that these subsets share the same event and censored ratio. 

\subsection{Implementation details}

Non-image time-dependent data is represented by a matrix of size $D \times 2W$, where $D$ is the cut-off day, i.e. the maximum number of days of data the model could use in this event prediction, and $W$ is the total number of types of lab results available. For each vector of length $2W$ in this matrix, the first $W$ values $x_1, \dots, x_W$ are the lab results re-scaled to 0-1 range, and the remaining $W$ values $y_1, \dots, y_W$ are indicators. $y_i=1$ if the corresponding $x_i$ is a lab measurement available on that day, and $0$ otherwise.

X-ray images of the same patient are downsized and stacked, resulting in a tensor of size $224 \times 224 \times O$, where $O$ is the number of patient X-rays available for this event. 

To predict the risk function $r(x_j)$ for patient $j$, the image data is fed to a convLSTM (kernel size $7 \times 7$, 2 hidden layers size $32$ and $16$) recurrently to extract image features. The extracted features are further downsized by a 2D Adaptive Average Pooling layer ($2\times2$) following a fully connected layer (size 64). Non-image time-dependent inputs are first mapped to an embedding space (embedding size 15), then fed recurrently to LSTM to extract the respective features. Patient demographics, which are all discrete categorical values, are first mapped to their corresponding embedding space (embedding size 2), then concatenated altogether with the extracted features from convLSTM and LSTM branches. The concatenated vector goes through three fully connected layers (size 32, 16 and 1) to predict the final outcome.

During training, we use Adam \cite{AdamICLR2015} with $lr = 1e-3$, with a batch size of 40. We divide the training phase into 30 epochs, each epoch consists of 20 randomly sampled mini-batches. With limited GPU memory available and to reduce computational costs, we further sample the X-ray data and limit the number of X-rays $k$ to $k = 4$ per patient, and use zero padding when patients have less than $k$ valid X-rays.

In the validation phase, for each epoch we choose the mini-batch that the model has the lowest concordance error on as our baseline hazard, and compute the concordance error on the validation set. The epoch with the lowest concordance error and its best-performing mini-batch on validation set will be used to compete against other baseline models on the testing set.

\subsection{Baseline details}

We compare our proposed architecture with 8 different models, including parametric and semi-parametric models, standard and non-linear CoxPH models, and other popular machine learning survival methods: 
\begin{itemize}
    \item \textbf{Non-Image Input Models: } Weibull, Gompertz \cite{kleinbaum2012parametric}, Survival Forest \cite{ishwaran2008random}, Survival SVM \cite{van2007support}, the standard Cox Proportional Hazard model (CoxPH) \cite{cox1972regression}, the non-linear CoxPH model (DeepSurv)\cite{katzman2018deepsurv}
    \item \textbf{Image Input Models: } Deep Convolutional Neural Network for Survival Analysis (DeepConvSurv) \cite{zhu2016deep}
\end{itemize}

Recall that for our model, the non-image time-dependent input are formatted to matrix of size $D \times 2W$. For the non-image baselines, which take 1D input, we concatenate the row vectors in the matrix along the day axis, and further combine it with time-fixed data to obtain a long vector representation of the patient data. 

For DeepSurv \cite{katzman2018deepsurv}, we used a model with 2 fully connected layers (size 128, 64) with ReLu activation. 
For DeepConvSurv \cite{zhu2016deep}, we implemented a network very similar to the original, which consists of three convolution layers (conv1: kernel size $7 \times 7$, stride 3, channel 32; conv2: kernel size $5 \times 5$, stride 2, channel 32; conv3: kernel size $3 \times 3$, stride 2, channel 32), following a single FC layer. We also added max pooling and 2D dropout to the network to improve model performance. Similar to our own architecture, we train DeepConvSurv using randomly selected images for each patient, and experiment on both the case where the loss is computed over the entire training set, and the case where mini-batch gradient descent is used.

To investigate the impact of images on prediction, we provide two baselines of our own: 1) a model where the convLSTM branch is removed and predicts only from non-image data, and 2) an ``complementary" model where the LSTM branch is removed and we predict solely from image-data.
In these two baseline models, we also use mini-batch gradient descent with batch size of 40.

\section{Results}
\label{sec:data}

\subsection{Performance comparisons}
\label{sec:comparison}

We evaluate model performance using time-dependent concordance error discussed in section~\ref{sec:eval-and-infer}. 
Results of the comparisons are in Table~\ref{tab:main}. 
Additional experiments and ablation studies are included in the supplemental material.

\begin{table*}[ht!]
    \centering
\begin{tabular}{|c | c c c c c|}
\hline
   Method & Admit & ICU Admit & ICU Discharge & Discharge & Death \\
\hline\hline
CoxPH & 0.434 & 0.371 & 0.486  & 0.240\cellcolor{Magic Mint}  & 0.358 \\
Weibull & 0.319 & 0.372 & 0.478   & 0.437  & 0.477  \\
Gempertz & 0.382 & 0.362  &  0.413 \cellcolor{Magic Mint}  & 0.272 & 0.360\\
DeepSurv & 0.316 & 0.358  &  0.454  & 0.398 & 0.373 \\
Survival+SVM & 0.333 & 0.365  &  0.464  & 0.347 & 0.316 \\
Survival+Random Forest & 0.337 & 0.309 \cellcolor{Aero Blue}  &  0.435\cellcolor{Aero Blue}  & 0.427 & 0.282\cellcolor{Aero Blue} \\
\hline
DeepConvSurv (GD) & 0.403 & 0.417  &  0.468  & 0.356 & 0.351 \\
DeepConvSurv (mini-batch) & 0.419 & 0.415  &  0.478  & 0.419 & 0.435 \\
\hline
Image-Only (ours) & 0.238\cellcolor{Aero Blue} & 0.408 &  0.459  & 0.350 & 0.401\\
Non-Image (ours) & 0.247\cellcolor{Magic Mint} & 0.265\cellcolor{Magic Mint}  &  0.439 & 0.262\cellcolor{Aero Blue} & 0.278\cellcolor{Magic Mint} \\
Image+Non-Image (ours) & \cellcolor{Pearl Aqua}\textbf{0.198} & \textbf{0.241}\cellcolor{Pearl Aqua}  & \cellcolor{Pearl Aqua}\textbf{0.385} & \cellcolor{Pearl Aqua}\textbf{0.229} & \cellcolor{Pearl Aqua}\textbf{0.246} \\
\hline
\end{tabular}
    \caption{Concordance error for time-to-event predictions. Lower is better, colors encode the best performing 3 methods for each event. DeepConvSurv with Gradient Descent (GD) computes its loss function over the entire patient group, while DeepConvSurv with mini-batch Stochastic Gradient Descent (mini-batch) adopts the same sampling strategy as our model described in section~\ref{sec:mini-batch}.}
    \label{tab:main}
    \vspace{-0.5cm}
\end{table*}

In our experiments, non-linear CoxPH models (DeepSurv and ours) and Random Survival Forest almost always outperform parametric models.
This is expected since Cox models make no distributional assumptions about the data. 
The performances of DeepConvSurv and our Image-Only baseline are also noteworthy: they demonstrates that image alone (without any lab values or demographics) provides useful information for prediction. Using multiple images further improves the performances. 

Our recurrent baseline model, which captures time-dependency relations, achieves competitive performance across all 5 events even without images. 
Adding our ConvLSTM branch that processes time-dependent images has further improved the predictions by an average of $2.8\%$, with a substantial improvement of $4.9\%$ on Hospitalization (Admit) and $5.4\%$ on ICU Discharge.

\begin{figure}
    \hspace*{-.3cm}
    \includegraphics[width=1.1\linewidth,trim=0 0 1in 0,clip]{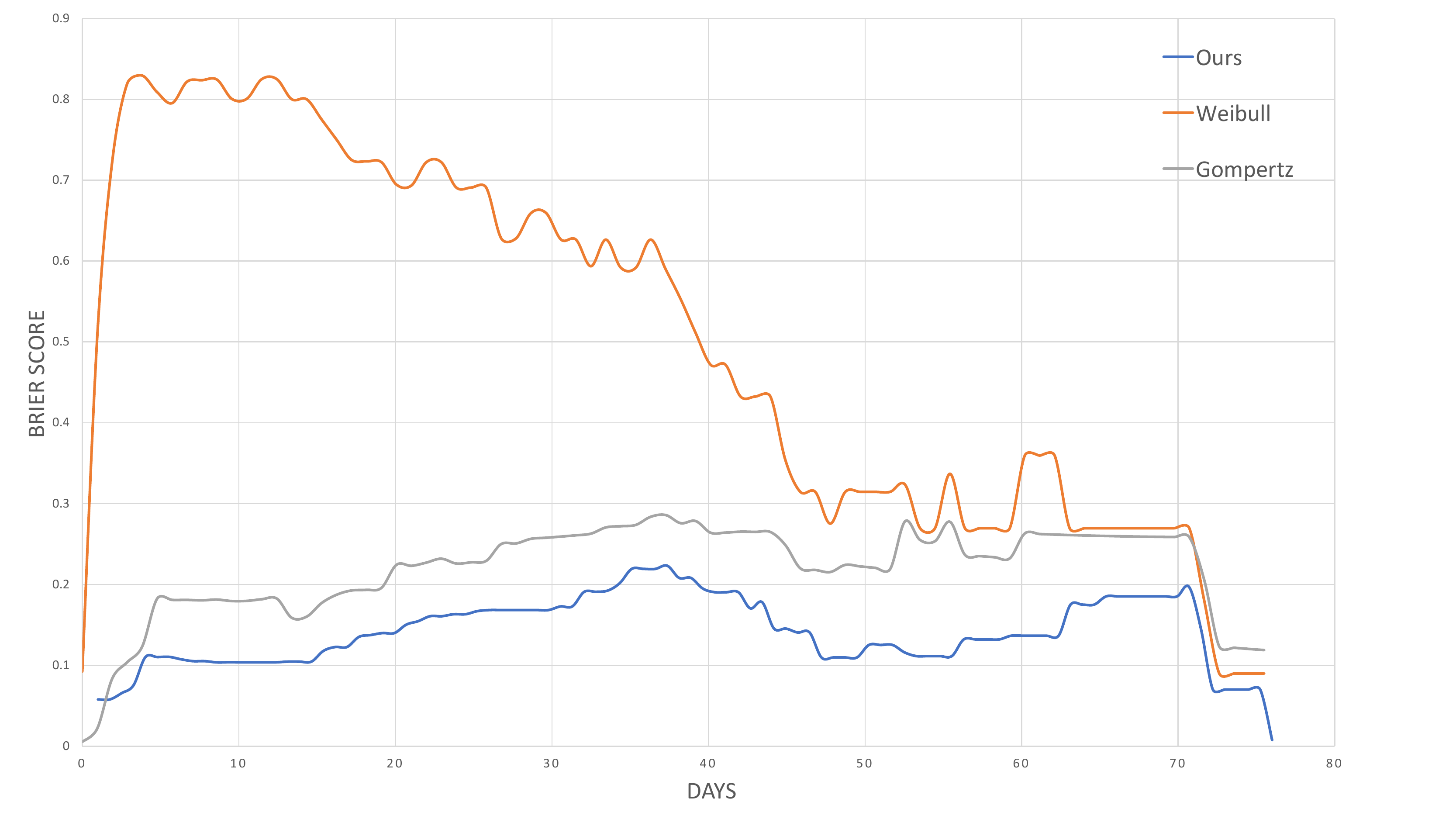}
    \caption{Brier score comparison against selected standard techniques, lower is better}\label{fig:Brier}
    \vspace*{-.5cm}
\end{figure}

To better illustrate the comparison, in figure ~\ref{fig:Brier} we show the time-dependent Brier score \cite{schoop2008quantifying} for our method and two standard time-to-event prediction techniques. This is an extension of the Brier score \cite{graf1999assessment} for a specific time horizon.

Our experiments demonstrate the effectiveness of our recurrent architecture on non-linear, longitudinal data. 
We also show that incorporating multiple, time-dependent imaging studies significantly improves time-to-event predictions.

\subsection{Understanding the dataset and model}
We conduct several experiments to provide insights into what our models learned.

\textbf{Feature Importance Test: } While our imaging studies were anonymized, fields describing individual scanners were preserved. This provides some ways to check if our model is simply learning an association between certain scanners and disease severity 
(for example, sicker patients might be in certain areas of the hospital, and routed to the nearest x-ray machine) . 
To test whether our model is learning simple associations between scanners and events, we conducted permutation-based feature importance tests (using survival random forest) to measure performance drops if certain features are removed.
Specifically, we add scanner IDs of patient x-rays as one of the features for event prediction, and measure the average concordance error increase when the relationship between the feature and survival time is removed by random shuffling. We
found that for Hospital Admission and ICU Admission, Scanner ID has an average concordance error increase of $0.0020 \pm 0.0044$ and $0.0060 \pm 0.0703$ respectively, and $0.00 \pm 0.000$ on the ICU Discharge, Discharge and Death events. 
Compared with other features such as Age ($0.0682 \pm 0.0542$ on ICU Discharge), the permutation tests results on Scanner ID suggest little correlation between scanner types and events.

\textbf{Black Box Test: }Recent papers, such as \cite{maguolo2020critic}, 
have identified flaws in the evaluation of deep learning models for COVID-19. Deep learning models were able to utilize non-medical information in the X-rays like artifacts or watermarks in order to differentiate between patients with COVID and those without. In order to demonstrate that our dataset and model do not exhibit this flaw, we performed similar experiments to \cite{maguolo2020critic}, where we obscured various portions of the images with black boxes and retrained the model. This should  determine what part of the images are important to the model. If our technique is relying on incidental properties of the images, such as identifiable artifacts from individual scanners, we would expect to see better prediction accuracy from an image and labs model over a labs-only model even when most of the image is obscured.

\begin{table*}
  \centering
  \vspace*{.3in}
  \includegraphics[scale=.5,trim= 0 200 0 50]{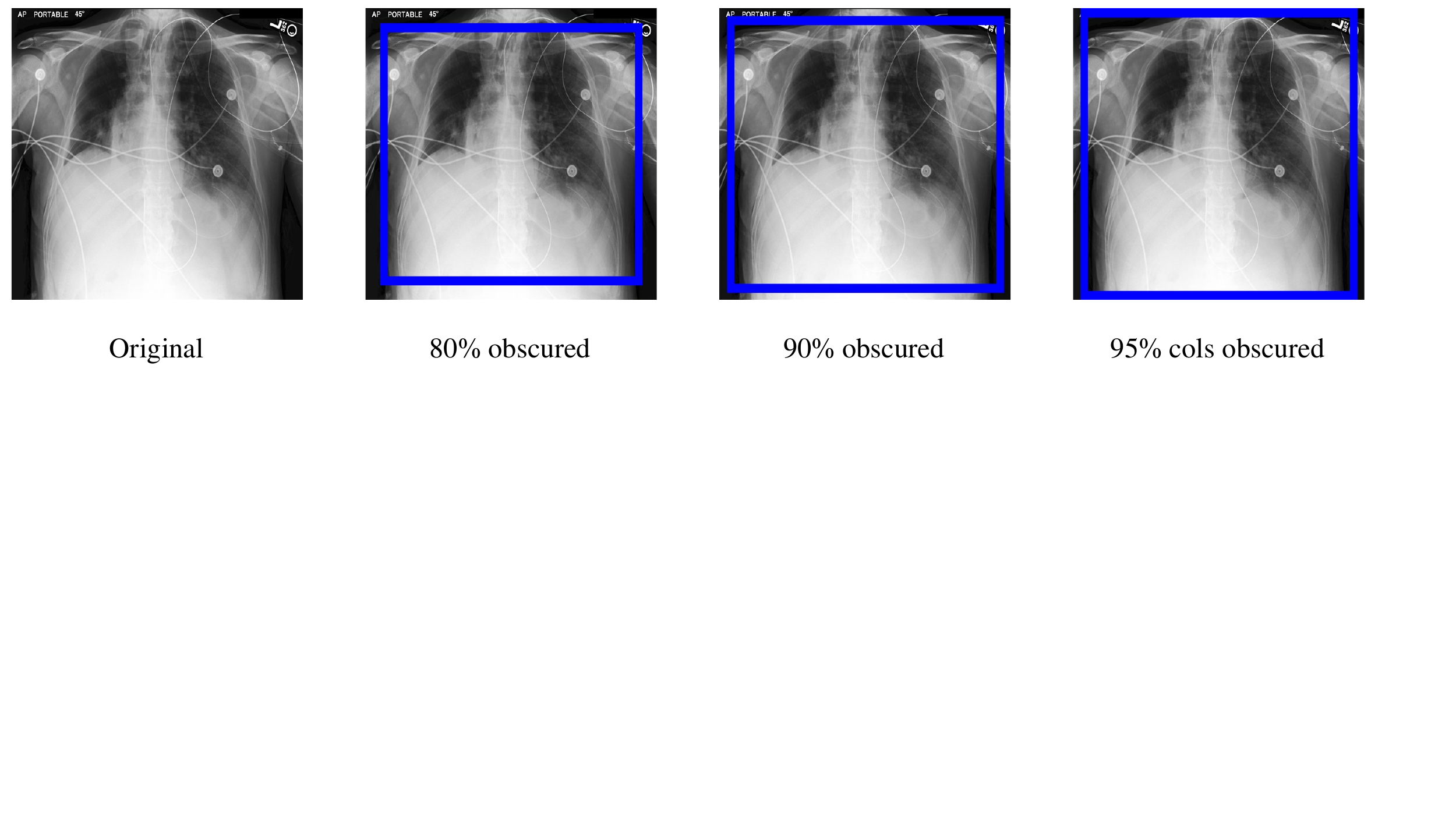}
    \begin{tabular}{|l||c|c|c|c|c|c|}
    \hline
    Event & \multicolumn{1}{p{4.215em}|}{\centering Image + labs (ours)} & \multicolumn{1}{p{4.215em}|}{\centering 80\% obscured} & \multicolumn{1}{p{4.215em}|}{\centering 90\% obscured} & \multicolumn{1}{p{4.215em}|}{\centering 95\% cols obscured} &  \multicolumn{1}{p{4.215em}|}{\centering 100\% \mbox{obscured}}  & \multicolumn{1}{p{4.215em}|}{\centering Labs only (ours)} \\
    \hline
    Admit Date & \textbf{0.198} & 0.250 & 0.257 & {0.287} & 0.268 & 0.247 \\
    \hline
    ICU Admit & \textbf{0.241} & 0.262 & 0.271 & {0.262} & 0.264 & 0.265 \\
    \hline
    ICU Discharge & \textbf{0.335} & 0.413 & 0.437 & {0.431} & 0.437 & 0.439 \\
    \hline
    Discharge & \textbf{0.229} & 0.272 & 0.268 & {0.276} & 0.264 & 0.262 \\
    \hline
    Death & \textbf{0.254} & 0.296 & 0.274 & {0.271} & {0.278} & 0.278 \\
    \hline
    \end{tabular}%
  \caption{Ablation on obscuring parts of the input images. The images on the top provide a visual example for each input type; note that the portion inside the blue box is set to zero. 
  (Original x-ray image taken from a publicly available dataset.)
  The table shows concordance error for time-to-event predictions (lower is better). If our model was learning coincidental features outside the patient's body, such as scanner artifacts, we would expect that obscuring the center of the image would result not significantly degrade performance, which is not the case. 
  \label{tab:data}}
  \label{tab:addlabel}%
\end{table*}%

Results are shown in table~\ref{tab:data}. 
The first and last columns give our results with images included (first column) and non-image data only (last column).
We used several different sized boxes and zeroed out the portions of the image within them.
As a sanity check we obscured all of the image (``100\% obscured'').
We also obscured large portions of the image (``80\% obscured'', ``90\% obscured'', the percentages indicate how much of the image was obscured starting at the center).
Note that sometimes clues concerning illness severity (e.g. tubes) are located at the top, and thus visible in the 80\% obscure and 90\% obscure cases. Therefore, we also included a ``95\% cols obscured"  case, where 95\% of the center columns are   removed. 
The boxes are shown overlayed on a publicly available chest x-ray above table~\ref{tab:data}. 

The experimental results suggest that our models are not learning coincidental features such as scanner artifacts or watermarks. We also observe that there are some small improvements when our models use 80\% obscured images, and there is a natural explanation: unless the image is almost entirely obscured, various lines and sensors are easily discernible even outside the lungs, and these possibly provide an indication of disease severity.

\textbf{Number of Input Images Ablation:} We also perform an ablation on the number of images used ($k$) when training and evaluating the model. We find that the model improves as the number of images increases for the majority of tasks (admission, ICU discharge, and discharge). Specifically, we observe that $k=2$ provides a 10-20\% improvement in concordance error over $k=1$; for example, going from $k=2$ to $k=1$ leads to a decrease in concordance error of $0.072$ in admission; $0.006$ in discharge; and $0.048$ in ICU discharge. The remaining tasks show little to no improvement from a larger number of images. However, an analysis of the dataset statistics show that these tasks tend to have fewer images per patient; this means that our training set also has fewer examples of patients with more data which may leads to problems with generalization. We include more details on this ablation in the supplemental material.

\begin{figure}
    \includegraphics[width = \linewidth, scale=0.8]{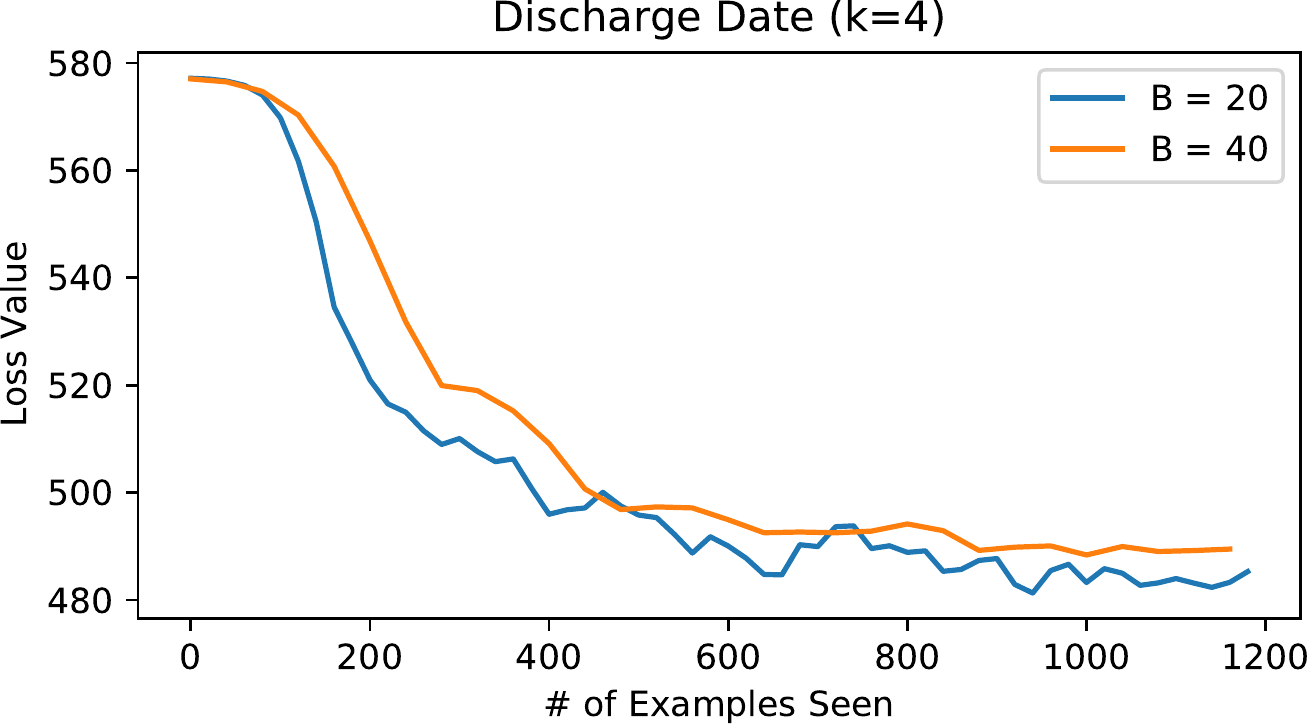}
    \caption{Validation error curves for models trained with $k = 4$ images per patients and mini-batch size $B = 20$ or $40$ respectively. The convergence properties do not seem to be affected by mini-batch size.}\label{fig:val_curves}
    \vspace{-0.5cm}
\end{figure}

\textbf{Mini-batch Size Ablation:} We also investigated the effects of mini-batch size $B$, and observed that mini-batch size does not affect the model performance for sufficiently large values of $B$ ($B\geq 20$).  Particularly small mini-batches ($B=10$) contains very few events, which results in performance drops. However, for $B\geq 20$, the models tend to converge to the same point. In figure~\ref{fig:val_curves}, we show an example of the validation error curves of discharge events with respect to the number of examples seen for the $k = 4$ case, where $B=20$ and $B=40$. From the figure, we see that the size of the mini-batch does not seem to affect convergence. 

\section{Conclusions}

We describe a deep learning approach that incorporates multiple, time-fixed data, longitudinal non-image data and longitudinal images into time-to-event analysis. 
Our technique accurately predicts the probability of experiencing an event in the presence of right-censored data.

We used a large COVID-19 dataset containing longitudinal imaging
and non-imaging information.
While this dataset contains valuable information to predict the occurrence of clinical events, there is some risk of selection bias. 
For instance, we only included in our analysis patients with multiple imaging studies over time. 
Multiple images are usually performed when a patient is sicker, for example to confirm central line placement. 
This selection could lead to a sample that is not representative of the COVID-19 hospitalized population. 
Selection bias is a common problem in machine learning \cite{cortes2008sample}, statistics \cite{whittemore1978collapsibility}, and epidemiology \cite{robins2001data}; as a result,  a number of techniques have been developed to correct it \cite{cortes2008sample}. 

We have demonstrated that neural networks can be used to explicitly support transitions between competing events and be used to predict the transition-specific risk (e.g., the risk of transition from hospitalization to ICU admission) for a particular set of patient features.
While our focus is on COVID-19, the techniques we propose should be generally applicable to a wide class of serious diseases where imaging can improve the prediction of patient outcomes.

\vspace*{.2cm}
\textbf{Acknowledgements} We thank Joshua Geleris MD and George Shih MD for their help understanding various clinical issues. We received considerable help acquiring the dataset from Martin Prince MD PhD (PI on the IRB), Benjamin Cobb, Sadjad Riyahi MD, and Evan Sholle. This research was supported by a gift from Sensetime.
\clearpage
\newpage
\bibliographystyle{unsrt}
\bibliography{RDZ-vision-bib,R21-bib,cvpr21}

\end{document}

\title{Supplementary material for submission \#7091:\\
{{Deep survival analysis with longitudinal X-rays for COVID-19}}}  %
\date{}

\maketitle

\thispagestyle{empty}

\section{Parameter sensitivity study}
In the main paper, we investigated both the number of images and the size of mini batch, and concluded that 1) the model improves as the number of images increases for the majority of tasks (admission, ICU discharge, and discharge) and 2) mini-batch size does not affect the model performance for sufficiently large values of $B$ ($B\geq 20$).

Table~\ref{tab:parametersearch} summarizes the results.

\begin{table}[ht!]
{\small
\centering
\begin{tabular}{|c| c c c c|}
\hline
Admission    & $k=8$ & $k=4$ & $k=2$ & $k=1$\\
\hline\hline
$B = 20$ & 0.195 & 0.191  & 0.180\cellcolor{Magic Mint}    & 0.272 \\
$B = 40$ & - & \textbf{0.171}\cellcolor{Pearl Aqua}   & 0.201  & 0.303 \\
$B = 80$ & - & - & 0.198  & 0.232 \\
$B = 142$ (All) & -  & -   &  0.193 \cellcolor{Aero Blue} &  0.252 \\
\hline\hline
ICU Admission    & $k=8$ & $k=4$ & $k=2$ & $k=1$\\
\hline \hline
$B = 20$ & 0.259 & 0.258  & 0.258  & 0.285 \\
$B = 40$ & - & \textbf{0.234}  \cellcolor{Pearl Aqua}  & 0.241 \cellcolor{Aero Blue}  & 0.287 \\
$B = 80$ & - & - & 0.254  & 0.278 \\
$B = 160$ & -  & -   &  0.237 \cellcolor{Magic Mint} &  0.277 \\
\hline
ICU Discharge    & $k=8$ & $k=4$ & $k=2$ & $k=1$\\
\hline \hline
$B = 20$ & 0.345 & 0.331  & \textbf{0.306} \cellcolor{Pearl Aqua} & 0.427 \\
$B = 40$ & - & 0.335    & 0.337  & 0.366 \\
$B = 80$ & - & - & 0.327\cellcolor{Magic Mint}  & 0.341 \\
$B = 160$ & -  & -   &  0.330 \cellcolor{Aero Blue} &  0.358 \\
\hline
Discharge    & $k=8$ & $k=4$ & $k=2$ & $k=1$\\
\hline \hline
$B = 20$ & 0.230 & 0.231  & 0.229  & 0.240 \\
$B = 40$ & - & 0.229  \cellcolor{Magic Mint}  & 0.232  & 0.233 \\
$B = 80$ & - & - & 0.231\cellcolor{Aero Blue}  & 0.235 \\
$B = 160$ & -  & -   &  \textbf{0.228} \cellcolor{Pearl Aqua} &  0.238 \\
\hline
Death   & $k=8$ & $k=4$ & $k=2$ & $k=1$\\
\hline \hline
$B = 20$ & 0.268 & 0.271  & 0.279  & 0.262 \\
$B = 40$ & - & 0.247 & 0.346  & \cellcolor{Aero Blue} 0.239 \\
$B = 80$ & - & - & \textbf{0.232}\cellcolor{Pearl Aqua}  & 0.249 \\
$B = 160$ & -  & -   &  0.238 \cellcolor{Magic Mint} &  \cellcolor{Aero Blue} 0.239 \\
\hline
\end{tabular}
    \caption{Parameter sensitivity study on mini-batch size $B$ and maximum number of x-rays  per patient $k$ for Discharge and ICU Discharge Predictions. Performance is measured by concordance error, lower is better. The experiments are conducted on hospitalization (admission) prediction. When $B = 142$, all available patients from the training set are included in the mini-batch.}
    \label{tab:parametersearch}
    \vspace{-0.5cm}
}
\end{table}

\newpage
\section{Non-Image Data Only Comparisons}
We conduct an ablation test on data filtering. As mentioned in our data pipeline, to compare with models that only take images as input, we removed patients that do no have any x-rays taken during the target interval. As a sanity check, the study below uses unfiltered data and compare non-image only models. 

Table ~\ref{tab:main} compare results of ICU Admit, ICU Discharge and (Hospital) Discharge predictions, where there are substantial population changes when we applied the filter. We keep the same event-censor ratio, and split the data randomly into train, eval, and test by $60:20:20$.

The results provide evidence that our technique works well even in the absence of images, independent of the data filtering.

\begin{table*}
    \centering
\begin{tabular}{|c|c|c|c|c|}
\hline
   & ICU Admit & ICU Discharge & Discharge \\
\hline\hline
Event \# & 420 & 252  & 1197 \\
Censored \# & 1099 & 173  &  333 \\
\hline
\end{tabular}
    \caption{Details on the data distribution for each type of events for unfiltered data ablation test.}
    \label{tab:data_distribution}
\vspace{0.5cm}
\end{table*}

\begin{table*}
    \centering
\begin{tabular}{|c |c c c|}
\hline
   Method & ICU Admit & ICU Discharge & Discharge\\
\hline\hline
CoxPH (Full) & 0.414 & 0.373 & 0.273\\
Weibull (Full) & 0.331 \cellcolor{yellow} & 0.324 \cellcolor{orange} &  0.264\\
Gempertz (Full) & 0.345 & 0.326  &  0.185 \cellcolor{orange} \\
DeepSurv & 0.358 & 0.337  &  0.186 \\
Survival+SVM & 0.355 & 0.397  &  0.215\\
Survival+Random Forest & 0.286\cellcolor{red} & 0.307 \cellcolor{yellow} &  0.206 \cellcolor{yellow}\\
\hline
Ours & 0.302 \cellcolor{orange}& 0.238 \cellcolor{red} & 0.173 \cellcolor{red}\\
\hline
\end{tabular}
    \caption{Concordance error for time-to-event predictions. Lower is better, colors encode the best performing 3 methods for each event. Full models take input concatenated across all available days, while cross-sectional (CS) models only take time-dependent data from the baseline date, which satisfies the PH assumption.}
    \label{tab:main}
    \vspace{-0.5cm}
\end{table*}

\begin{table}
    \centering
\begin{tabular}{|c | c |c |}
\hline
Type & Abbreviation &Unit \\
\hline \hline
Alanine Transferase & ALT & $U/L$\\
Glucose Level & - & $mg/dL$\\
Albumin Level & Alb & $g/dL$\\
Aspartase Transaminase & AST & $U/L$\\
Ala Aminotransferase  &ALA & $U/L$\\
Aspartate Aminotransferase & ASA & $U/L$ \\
Creatinine & Cr & $mg/dL$\\
Lymphocyte & - & $cells/uL$ \\
Creatine Kinase & CK & $U/L$ \\
Lactate Dehydrogenase & LDH & $U/L$ \\
Ferritin Level & - & $ng/mL$ \\
B-Type Natriuretic Peptide & BNP & $pg/mL$ \\
Troponin I & Trop & $ng/mL$ \\
Sedimentation Rate & ESR & $mm/hr$  \\
C-Reactive Protein & CRP & $mg/dL$\\
D-Dimer, & - & $ng/mL$ \\
Interleukin-6  & IL-6 & $pg/mL$ \\
Platelets & - & $\times10^3/uL$\\
Thromboplastin Time & PTT & $s$ \\
\hline
vs\_hr\_hr & - & -\\
xp\_resp\_spo2 & - & - \\
xp\_resp\_rate\_pt & - & - \\
vs\_bp\_noninvasive (s) & - & -\\
vs\_bp\_noninvasive (d) & - & - \\
vs\_bp\_noninvasive (m) & - & -\\
HCO3 (Arterial) & - & $mmol/L$ \\
pO2 (Arterial) & - & $mmHg$\\
pCO2 (Arterial) & - & $mmHg$\\
pH (Arterial) & - & -\\
\hline

\end{tabular}
    \caption{List of time-dependent features used in our technique.}
    \label{tab:params}
    \vspace{-0.5cm}

\end{table}

\begin{table*}
    \centering
\begin{tabular}{|c|c|c|c|c|c|}
\hline
   & Admit & ICU Admit & ICU Discharge & Discharge & Death  \\
\hline\hline
Mean & 2.72 & 1.78  & 6.65 & 2.88 & 4.61 \\
\hline
Std &  3.39 & 1.11  & 4.61 & 3.39 & 4.00 \\
\hline
\end{tabular}
    \caption{Details on the number of images per patient within data range}
    \label{tab:data_distribution}
\vspace{0.5cm}
\end{table*}

{\small
}